\documentclass[aps,prd,showpacs]{revtex4} 
\usepackage[dvips]{graphicx}
\usepackage{wrapfig}
\usepackage{bm}
\begin{document}    
\title{Deviation from  tri-bimaximal neutrino  mixing
in  $A_4$ flavor symmetry}

\author{Mizue Honda}
\affiliation{Graduate School of Science and Technology, 
Niigata University, Niigata 950-2181, Japan}
\author{Morimitsu Tanimoto}
\affiliation{Department of Physics, Niigata University, Niigata 950-2181,
Japan}

\date{\today}
 
\begin{abstract}
  
The tri-bimaximal mixing  is a good approximation for the 
present data of  neutrino mixing angles.  
The deviation from the tri-bimaximal mixing is discussed numerically
 in the framework of the  $A_4$ model.
 Values of $\tan^2\theta_{12}$, $\sin^2 2\theta_{23}$ and
 $|U_{e3}|$  deviate from the tri-bimaximal mixing 
due to the  corrections of the vacuum alignment of flavon fields.
It is remarked that  $\sin^2 2\theta_{23}$ deviates  scarcely from $1$
while  $\sin^2 \theta_{12}$ can  deviate from $1/3$ considerably
and  $\sin\theta_{13}$ could be near the present experimental upper bound.
  The CP violating measure $J_{CP}$ and 
 the effective Majorana neutrino mass $\langle m_{ee} \rangle$ 
 are also discussed.
\end{abstract}


\maketitle

\section{Introduction}

Neutrino experimental  data provide us  an important clue to find an origin 
of the observed hierarchies in mass matrices for quarks and leptons.
Recent experiments of the neutrino
 oscillation go into the new  phase  of precise  determination of
 mixing angles and mass squared  differences \cite{maltoni,fogli}. 
Those indicate the  tri-bimaximal mixing  for three flavors 
 in the lepton sector \cite{HPS}. 
Therefore, it is very important
to find a natural model that leads to this mixing pattern with good accuracy.

 Flavor symmetry is expected to explain  the mass spectrum  and
the mixing matrix of quarks and leptons. 
Especially,  some predictive models with discrete flavor
symmetries have been explored by many authors~\cite{S3}-\cite{Alta2}.
Among them,
the  interesting models to give the tri-bimaximal mixing
 are  based on the the non-Abelian finite group $A_4$.
Since the original papers \cite{A4} on the application of the 
non-Abelian discrete symmetry $A_4$ to quark and lepton families, 
much progress has been made in understanding the  tri-bimaximal 
mixing for neutrinos in a number of specific models \cite{A41}-\cite{Alta2}.
 Therefore, it is important to  test  the  $A_4$ model experimentally.
 
 We present the comprehensive analyses of 
 the deviation from the tri-bimaximal mixing  
 in the framework of  the $A_4$ model,
where the tri-bimaximal mixing is realized in  the vacuum alignment
of the flavon fields \cite{Alta1,Alta2}.
Since the vacuum alignment is corrected  by  higher-dimensional operators,
the deviation from the  tri-bimaximal mixing  is  predicted numerically.

It is found  that  $\sin^2 2\theta_{23}$  deviates  scarcely  from $1$
while  $\sin^2 \theta_{12}$ can  deviate from $1/3$ considerably
and  $\sin\theta_{13}$ could be near the present experimental upper bound.
  The CP violating measure $J_{CP}$ and 
 the effective Majorana neutrino mass $\langle m_{ee} \rangle$ 
 are also predicted.

The paper is organized as follows:
 we present the framework of the model in Sec. II, and
 discuss the deviation from the  tri-bimaximal mixing of neutrino flavors
in  Sec. III.
In Sec.IV, the numerical results are presented.
Section V is devoted to the summary.
The useful relations among parameters are given in Appendix.
 
\section{Framework of the $A_4$ Model}
 The tri-bimaximal mixing pattern  is a good approximation for the 
present data of  neutrino mixing angles.  Therefore, it is very important
to find a natural model that leads to this mixing pattern with good accuracy.
The  interesting models to give the tri-bimaximal mixing
 are based on  the non-Abelian finite group  
$A_4$, in which there  are twelve group elements and
 four irreducible representations: $\bf 1$, $\bf 1'$, $\bf 1''$ and  $\bf 3$.
Under the $A_4$ symmetry, the left-handed lepton doublets $\ell_L$ 
are assumed to transform as  $\bf 3$, 
the right-handed charge lepton singlets $e^c$,  $\mu^c$ and  $\tau^c$ as   
$\bf 1$, $\bf 1'$, $\bf 1''$, respectively.
The flavor symmetry is spontaneously broken 
by two real   ${\bf 3}'s$, $\varphi$, $\varphi'$, and by three  real singlets,
 $\xi(\bf 1)$, $\xi'(\bf 1')$  and $\xi''(\bf 1'')$,
which are $SU(2)_L$ gauge singlets.

 The relevant  Yukawa  couplings of leptons are given as follows:
\begin{eqnarray}
{\cal L}_Y&=& \frac{y_e}{\Lambda} e^c(\varphi\ell_L)h_d
+ \frac{y_\mu}{\Lambda}  \mu^c(\varphi\ell_L)''h_d
+ \frac{y_\tau}{\Lambda}\tau^c(\varphi\ell_L)' h_d
+  \frac{x_a}{\Lambda^2} \xi (\ell_L h_u \ell_L h_u) + \nonumber\\
&&
+  \frac{x_b}{\Lambda^2}   \xi'' (\ell_L h_u \ell_L h_u)'
+  \frac{x_c}{\Lambda^2}  \xi' (\ell_L h_u \ell_L h_u)'' 
+  \frac{x}{\Lambda^2} (\varphi' \ell_L  h_u \ell_L h_u)
+ h.c.  \ ,
\label{La}
\end{eqnarray}
\noindent
where $h_d$ and $h_u$ are   ordinary Higgs scalars, 
  $\Lambda$ is a cut-off scale, and $y_\alpha$, $x_i$ and $x$
are dimensionless coefficients with order one.
The effective Lagrangian without  $\xi'(\bf 1')$  and $\xi''(\bf 1'')$ 
was given by Altarelli and Feruglio \cite{Alta1}.
The  Lagrangian of Eq.(\ref{La}) is  a   general one  
to give the $A_4$ symmetric lepton mass matrices.
 The flavon fields $\varphi$,  $\varphi'$,  $\xi$,   $\xi'$ and   $\xi''$ 
develop the  vacuum expectation values along the directions:
\begin{eqnarray}
\langle \xi \rangle =u_a,  \quad
\langle \xi' \rangle =u_c, \quad
\langle \xi'' \rangle =u_b, \quad
\langle \varphi \rangle =(v_1,v_2,v_3), 
\quad\langle \varphi' \rangle =(v'_1,v'_2,v'_3) \ .
\end{eqnarray}

We take the three-dimensional unitary representation matrices of $A_4$
in Ref.\cite{A4}. Then,
if we put   $v_1=v_2=v_3=v$, the charged lepton mass matrix is given by
\begin{eqnarray}
M_{E}=  \sqrt{3} v_d \frac{v}{\Lambda}
U_0 \left(
\matrix{ y_e & 0 & 0 \cr 0 & y_\mu & 0 \cr 0 & 0 & y_\tau\cr}
\right)\ , 
 \label{charged}
\end{eqnarray}
\noindent where $v_d$ is the  vacuum expectation value of $h_d$ and
\begin{eqnarray}
U_{0}=
\frac{1}{\sqrt{3}}
\left(
\matrix{
1 & 1 & 1 \cr
1 & \omega & \omega^2 \cr
1 & \omega^2 & \omega \cr}
\right)\ ,  \qquad\qquad  \omega=\frac{-1+i\sqrt{3}}{2} \ .
 \label{U0}
\end{eqnarray}
\noindent
 On the other hand, the  left-handed Majorana mass matrix, which respects
the $A_4$ flavor symmetry, is written as
\begin{eqnarray}
M_{\nu}=
\left(
\matrix{
a+b+c & f & e \cr
f & a+\omega b+\omega^2 c & d \cr
e & d & a+\omega^2 b+\omega c \cr}
\right) \ ,
 \label{mnu}
\end{eqnarray}
\noindent 
 where 
\begin{eqnarray}
&&a=x_a u_a\frac{v_u^2}{\Lambda^2}\ , \quad
b=x_b u_b\frac{v_u^2}{\Lambda^2}\ , \quad
c=x_c u_c\frac{v_u^2}{\Lambda^2}\ , \nonumber\\
&&d=x v'_1 \frac{v_u^2}{\Lambda^2}\ , \quad\
e=x v'_2 \frac{v_u^2}{\Lambda^2}\ , \quad\
f=x v'_3 \frac{v_u^2}{\Lambda^2}\ ,
\end{eqnarray}
\noindent and  $v_u$ is the  vacuum expectation value of $h_u$.
\noindent
If $c=b$ and $e=f=0$ are taken in Eq.(\ref{mnu}),
 the neutrino mass matrix turns to
\begin{eqnarray}
M_{\nu}=
\left(
\matrix{
a+2 b & 0 & 0 \cr
0 & a -b & d \cr
0 & d & a - b \cr}
\right) \ .
\end{eqnarray}
\noindent
In the flavor diagonal basis of the charged lepton mass matrix,
the neutrino mass matrix is given as
\begin{eqnarray}
M_{\nu}^f=U_{0}^\dagger M_{\nu} U_{0}^*
=
\left(
\matrix{
a+\frac{2d}{3} & b-\frac{d}{3} & b-\frac{d}{3} \cr
b-\frac{d}{3} & b+\frac{2d}{3}  & a-\frac{d}{3} \cr
b-\frac{d}{3} & a-\frac{d}{3} & b+\frac{2d}{3} \cr}
\right) \  ,
\end{eqnarray}
\noindent which leads to the tri-bimaximal mixing of flavors
\begin{eqnarray}
U_{\rm tri-bi}=
\left(
\matrix{
\sqrt{\frac{2}{3}} & \frac{1}{\sqrt{3}} & 0 \cr
-\frac{1}{\sqrt{6}} & \frac{1}{\sqrt{3}} & -\frac{1}{\sqrt{2}} \cr
-\frac{1}{\sqrt{6}} & \frac{1}{\sqrt{3}} & \frac{1}{\sqrt{2}} \cr}
\right)\ ,
\end{eqnarray}
\noindent with three mass eigenvalues
\begin{eqnarray}
   a-b+d \ , \quad a+2b \ , \quad -a+b+d \ .
\end{eqnarray}
 Even if   $a=0$ or $b=c=0$ is taken, 
these mass eigenvalues give observed   two neutrino mass scales 
$\Delta m_{\rm atm}^2$ and $\Delta m_{\rm sol}^2$ 
although a moderate tuning of parameters is required. 
Actually, such models were presented in the previous works  
\cite{Ma20051,Alta1,Alta2}.

 It is important to discuss the origin of  conditions  $b=c$ and $e=f=0$, 
which realizes the tri-bimaximal mixing. 
If   $\xi'$ and $\xi''$ are decoupled in the Yukawa couplings,
  $b=c=0$ is obtained.
On the other hand, 
if the field $\varphi'$ develops  the vacuum expectation values 
 $\langle \varphi' \rangle =(v'_1, v'_2, v'_3)$ along the directions
 of  $v'_2=v'_3=0$, one can put   $e=f=0$.
 This vacuum alignment could be realized in the scalar potential
 with SUSY \cite{Alta1,Alta2}.

The condition  $b=c$ is not  expected 
  unless   $\xi'$ and $\xi''$ are decoupled.
Then,  the neutrino mixing  deviates  from the tri-bimaximal mixing.
On the other hand, the vacuum alignment  $v'_2=v'_3=0$ may be  modified.
Actually,  higher dimensional operators
contributing to the superpotential correct the  vacuum alignment
such as   $v'_2 \not= 0, \  v'_3 \not= 0$.
Moreover,  the  vacuum alignment 
 $\langle \varphi \rangle =(v, v, v)$ could be  also corrected
 through  higher dimensional operators. This effect contributes to
 the charged lepton mass matrix.
Therefore, the lepton flavor mixing 
 deviates from the tri-bimaximal one by   higher dimensional operators.
We discuss the pattern of this deviation quantitatively in this paper. 
 
\section{Deviation from the tri-bimaximal mixing}

Let us discuss the deviation from the tri-bimaximal mixing of  neutrinos.
There are three sources of the deviation as discussed 
  in the previous section. 
As far as  couplings  of  $\xi'$ and $\xi''$ are allowed
 in the Lagrangian of Eq.(\ref{La}),
  $b =c$ is not expected unless other symmetry is imposed on these couplings.
  One cannot control the magnitude of the $c/b$ ratio
 in the framework of the $A_4$ flavor symmetry.
 The  case of  $b\not = c$ has been discussed by Ma \cite{Ma20041}.
We will show  the numerical result, which is consistent with the result
 in  \cite{Ma20041}, in the next section.

On the other hand, 
if the field $\varphi'$ develops  the vacuum expectation values 
 $\langle \varphi' \rangle =(v'_1, v'_2, v'_3)$ along the directions
 of  $v'_2=v'_3=0$, one can put   $e=f=0$.
 The vacuum alignment was discussed in the scalar potential
 with SUSY \cite{Alta2}.  
It is found that this  vacuum alignment is spoiled by 
corrections of  higher-dimensional operators, which are 
suppressed by order $1/\Lambda$ \cite{Alta2}.
 The correction of the  vacuum alignment 
 $\langle \varphi \rangle =(v, v, v)$  is also investigated
in the scalar potential  with SUSY \cite{Alta2}.  
 The correction of the order of  $1/\Lambda$ may cause  the deviation
from the tri-bimaximal mixing through the charged lepton mass matrix.

There are 
direct corrections to masses of  charged leptons and 
 neutrinos  of order  $1/\Lambda^2$ and   $1/\Lambda^3$, respectively.
However, one can assign $Z_3$ charge to prevent new structures 
in  Eqs.(\ref{charged}) and (\ref{mnu})
such as $\ell_L,\ \varphi',\ \xi:\omega$ and  $e^c,\ \mu^c,\ \tau^c:\omega^2$
 \cite{Alta2}.
For example, the operator  $e^c(\varphi\varphi\ell_L) h_d$ 
does not correct  the charged mass matrix.
Operators
$(\varphi\varphi')'(\ell_L\ell_L)''h_u h_u$,
$(\varphi\varphi')''(\ell_L\ell_L)'h_u h_u$, and 
$\xi (\varphi\ell_L\ell_L)''h_u h_u$ contribute to
the neutrino mass matrix. The first and second operators give
corrections in  diagonal elements, which are absorbed in
 parameters  $b$ and $c$ of Eq.(\ref{mnu}), while 
the third one gives corrections in off diagonal elements, which are absorbed
in parameters  $d$, $e$ and $f$ of Eq.(\ref{mnu}).

At first step, let us  neglect the effect of  the correction 
on the  vacuum alignment  $\langle \varphi \rangle =(v, v, v)$.
Then, the deviation from the tri-bimaximal mixing comes from
 the neutrino sector.
  We take   parameters 
\begin{eqnarray}
c=b\ (1+\epsilon_{1})\ ,\quad  e=\epsilon_{2}\ d\ ,\quad  f=\epsilon_{3}\ d\ ,
\end{eqnarray}
\noindent
where non-zero $\epsilon_1$, $\epsilon_2$ and $\epsilon_3$ lead
to the deviation from the  tri-bimaximal mixing. 

 Then, the neutrino mass matrix is given as:
\begin{eqnarray}
M_{\nu}^f
=\left(
\matrix{
m_{11} & m_{12} & m_{13} \cr
m_{12} & m_{22} & m_{23} \cr
m_{13} & m_{23} & m_{33} \cr}
\right) \ ,
\end{eqnarray}
\noindent where
\begin{eqnarray}
m_{11}&\simeq&a+\frac{2d}{3}(1+\epsilon_{2}+\epsilon_{3}) ,\nonumber\\
m_{12}&\simeq&b-\frac{d}{6}\{2-(\epsilon_{2}+\epsilon_{3})-
           \sqrt{3}i(\epsilon_{2}-\epsilon_{3})\} ,\nonumber\\
m_{13}&\simeq&b(1+\epsilon_{1})-\frac{d}{6}\{2-(\epsilon_{2}+\epsilon_{3})+
           \sqrt{3}i(\epsilon_{2}-\epsilon_{3})\} ,\nonumber\\
m_{22}&\simeq&b(1+\epsilon_{1})+\frac{d}{3}\{2-(\epsilon_{2}+\epsilon_{3})+
           \sqrt{3}i(\epsilon_{2}-\epsilon_{3})\} ,\nonumber\\
m_{23}&\simeq&a-\frac{d}{3}(1+\epsilon_{2}+\epsilon_{3}) ,\nonumber\\
m_{33}&\simeq&b+\frac{d}{3}\{2-(\epsilon_{2}+\epsilon_{3})-
           \sqrt{3}i(\epsilon_{2}-\epsilon_{3})\} \ .
 \label{massf}
\end{eqnarray}

In the first order of the perturbation,
the neutrino masses $m_{i}$  are
\noindent
\begin{eqnarray}
m_{1}&=&a-b+d
         -\frac{b}{2}\epsilon_{1}
         -\frac{d}{2}(\epsilon_{2}+\epsilon_{3})
         +i \frac{\sqrt{3}}{6}d (\epsilon_{2}-\epsilon_{3})\ , \nonumber\\
m_{2}&=& a+2b
         +b\epsilon_{1}-d(\epsilon_{2}+\epsilon_{3})
         -i \frac{\sqrt{3}}{3}d (\epsilon_{2}-\epsilon_{3})\ ,\nonumber\\
m_{3}&=& -a+b+d-\frac{3b}{2}\epsilon_{1}
         +\frac{3d}{2}(\epsilon_{2}+\epsilon_{3})
         -i \frac{\sqrt{3}}{2}d (\epsilon_{2}-\epsilon_{3})\ ,
 \label{mass}
\end{eqnarray}
\noindent and the MNS matrix elements \cite{MNS} are
\begin{eqnarray}
U_{e1}&=&\sqrt{\frac{2}{3}}
               -\frac{d}{\sqrt{6}(3b-d)}(\epsilon_{2}+\epsilon_{3})\ ,
\nonumber\\
U_{e2}&=&\frac{1}{\sqrt{3}}
               +\frac{d}{\sqrt{3}(3b-d)}(\epsilon_{2}+\epsilon_{3})\ ,
\nonumber\\
U_{e3}&=-&\frac{b}{2\sqrt{2}(a-b)}\epsilon_{1}
               -\frac{d}{\sqrt{6}(2a+b-d)}i(\epsilon_{2}-\epsilon_{3})\ ,
\nonumber\\
U_{\mu 1}
         &=&-\frac{1}{\sqrt{6}}-\frac{3b}{4\sqrt{6}(a-b)}\epsilon_{1}
               -\frac{d}{\sqrt{6}(3b-d)}(\epsilon_{2}+\epsilon_{3})\ ,
\nonumber\\
U_{\mu 2}
         &=&\frac{1}{\sqrt{3}}
         -\frac{d}{2\sqrt{3}(3b-d)}(\epsilon_{2}+\epsilon_{3})
               -\frac{d}{\sqrt{2}(2a+b-d)}i(\epsilon_{2}-\epsilon_{3})\ ,
\nonumber\\
U_{\mu 3}
         &=&-\frac{1}{\sqrt{2}}+\frac{b}{4\sqrt{2}(a-b)}\epsilon_{1}
               +\frac{d}{\sqrt{6}(2a+b-d)}i(\epsilon_{2}-\epsilon_{3}),
\nonumber\\
U_{\tau 1}
         &=&-\frac{1}{\sqrt{6}}+\frac{3b}{4\sqrt{6}(a-b)}\epsilon_{1}
               -\frac{d}{\sqrt{6}(3b-d)}(\epsilon_{2}+\epsilon_{3})\ ,
\nonumber\\
U_{\tau 2}
         &=&\frac{1}{\sqrt{3}}
         -\frac{d}{2\sqrt{3}(3b-d)}(\epsilon_{2}+\epsilon_{3})
               +\frac{d}{\sqrt{2}(2a+b-d)}i(\epsilon_{2}-\epsilon_{3})\ ,
\nonumber\\
U_{\tau 3}
         &=&\frac{1}{\sqrt{2}}+\frac{b}{4\sqrt{2}(a-b)}\epsilon_{1}
               +\frac{d}{\sqrt{6}(2a+b-d)}i(\epsilon_{2}-\epsilon_{3})\ ,
 \label{MNS}
\end{eqnarray}
\noindent where all parameters are supposed to be real for simplicity.
It is  remarked that the $A_4$ phase $\omega$ turns to the CP violating phase
 if   $\epsilon_2\not =\epsilon_3$, that is , $e\not= f$ is realized.
The CP violation also  comes from phases in parameters $a$, $b$ and $d$.
 We will present numerical calculations by taking complex numbers
for these parameters.

In above  expressions, we have supposed the  vacuum alignment 
 $\langle \varphi \rangle =(v, v, v)$.
The vacuum alignment  may be  sizeably corrected
 through  higher dimensional operators.
Then, the charged lepton mass  matrix in Eq.(\ref{charged}) is not preserved.
 The charged lepton mass matrix is modified 
in terms of correction parameters $\epsilon^{ch}_1,\ \epsilon^{ch}_2$, 
which are defined
as $\langle \varphi \rangle= \{v, \ (1+\epsilon_1^{ch}) v, 
\ (1+\epsilon_2^{ch}) v \}$,
as follows:

\begin{eqnarray}
M_{E}&=&
   v_d \frac{v}{\Lambda} 
\left(
\matrix{
1 & 0 & 0 \cr
0 & 1+\epsilon_1^{ch} & 0 \cr
0 & 0 & 1+\epsilon_2^{ch} \cr}
\right)
\left(
\matrix{
1 & 1 & 1 \cr
1 & \omega & \omega^2 \cr
1 & \omega^2 & \omega \cr}
\right)
\left(
\matrix{ y_e & 0 & 0 \cr 0 & y_\mu & 0 \cr 0 & 0 & y_\tau\cr}
\right) \ .
\label{epch}
\end{eqnarray}
After a basis transformation of the left-handed charged lepton  
by the unitary matrix $U_0$  in Eq.(\ref{U0}), we get
$M_E'$ as
\begin{eqnarray}
 M_{E}' =  U_0^\dagger M_E =
v_d\frac{v}{\Lambda}\frac{1}{\sqrt{3}}
\left(
\matrix{
X y_e & Y y_\mu & Y^* y_\tau \cr
Y^* y_e & X  y_\mu & Y y_\tau \cr
Y y_e & Y^* y_\mu & X y_\tau \cr}
\right)\ ,
\end{eqnarray}
\noindent
where
\begin{eqnarray}
X&=&3+\epsilon_1^{ch}+\epsilon_2^{ch} \simeq 3\ , \nonumber\\
Y&=&\epsilon_1^{ch} \omega+\epsilon_2^{ch} \omega^2 
   = -\frac{1}{2}\{(\epsilon_2^{ch}+\epsilon_1^{ch})
                 +\sqrt{3}i(\epsilon_2^{ch}-\epsilon_1^{ch})\} \ ,\nonumber\\
Y^*&=&\epsilon_1^{ch} \omega^2+\epsilon_2^{ch} \omega 
     =-\frac{1}{2}\{(\epsilon_2^{ch}+\epsilon_1^{ch})
                           -\sqrt{3}i(\epsilon_2^{ch}-\epsilon_1^{ch})\}  \ .
\end{eqnarray}
Then, the left-handed mixing matrix of the charged lepton mass matrix
  is given by $U_0 U_{E}'$,  where $U_{E}'$ is given as
\begin{eqnarray}
U_{E}' \simeq
\left(
\matrix{
1 & \frac{Y}{X}& \frac{Y^*}{X}  \cr
-\frac{Y^*}{X}  & 1 & \frac{Y}{X}  \cr
-\frac{Y}{X}  & -\frac{Y^*}{X}  & 1 \cr}
\right) \ ,
\end{eqnarray}
\noindent
which is obtained by diagonalizing $M_E' M_E'^\dagger$.
Therefore, the deviation from the tri-bimaximal mixing due to
$\epsilon_1^{ch}$ and $\epsilon_2^{ch}$ is given as
\begin{eqnarray}
U_{MNS}=U_{E}^{'\dagger} \ U_{\rm tri-bi} \ ,
\end{eqnarray}
whose  relevant mixing elements are given as 
\begin{eqnarray}
U_{e1}&\simeq& \frac{1}{\sqrt 6}\left (2+\frac{Y+Y^*}{X}\right )
\simeq \sqrt{\frac{2}{3}}
\left [1-\frac{1}{6}({\epsilon_2^{ch}+\epsilon_1^{ch}})\right ]\ , \nonumber\\
U_{e2} &\simeq& \frac{1}{\sqrt 3}\left (1-\frac{Y+Y^*}{X}\right )\simeq
 \sqrt{\frac{1}{3}}\left [1+\frac{1}{3}({\epsilon_2^{ch}+\epsilon_1^{ch}})\right ]\ , \nonumber\\
U_{e3} &\simeq& \frac{1}{\sqrt 2}\frac{Y^*-Y}{X}
\simeq \frac{i}{\sqrt{6}}( \epsilon_2^{ch}-\epsilon_1^{ch})\ , \nonumber\\
U_{\mu 3} &\simeq& -\frac{1}{\sqrt 2}\left (1+\frac{Y^*}{X}\right ) 
\simeq  -\frac{1}{\sqrt 2} 
\left [1-\frac{1}{6}\left \{(\epsilon_2^{ch}+\epsilon_1^{ch})
-i \sqrt{3} (\epsilon_2^{ch}-\epsilon_1^{ch})\right \}\right ]\ .
\end{eqnarray}
It is noticed that a new CP violating phase appears
 due to  corrections of the vacuum alignment.

\section{Numerical results}
 Let us present  numerical results as for the deviation from
 the tri-bimaximal  mixing.
At first,  we discuss the case $a\not =0$,  $b\not =c$, $e=f= 0$
in  the neutrino mass matrix of Eq.(\ref{mnu})
 to see the effect of $\epsilon_1$, which denotes 
the deviation from  $b=c$. In this case,
we neglect the effect of   $\epsilon^{ch}_1,\ \epsilon^{ch}_2$ 
in the charged lepton mass matrix, therefore,
  the charged lepton mass matrix is given as in Eq.(\ref{charged}).

 Next, we discuss the case of 
 $a\not =0$, $b=c=0$, $e\not = 0$, $f\not =0$, in which 
 $b=c$ is guaranteed by vanishing Yukawa couplings as to $\xi'$ and $\xi''$.
This case leads to the normal mass  hierarchy of  neutrinos.
We also discuss the case of  the inverted mass hierarchy of neutrinos,
which could be realized in the case of $a=0$, $b = c$, $e\not = 0$, 
$f\not =0$.

The effect of the  correction  in the charged lepton sector
is  also discussed in the last subsection.

\subsection{$a\not =0$,  $b\not =c $,  $e=f= 0$}
The couplings  $\xi'$ and $\xi''$  generally lead to   $b\not =c $
 in the $A_4$ symmetry.
Then,  the neutrino mixing deviates  from the tri-bimaximal mixing
due to  $\epsilon_1$.
We suppose  the vacuum alignment  $v'_2=v'_3=0$, that is  $e=f= 0$
 in order to see the effect of  $b\not =c $ clearly.
This case has been studied analytically by Ma  at first \cite{Ma20041}.
Our numerical result is completely  consistent with the results 
in Ref. \cite{Ma20041} .

\begin{wrapfigure}{r}{7cm}
\begin{center}
\includegraphics[width=6 cm]{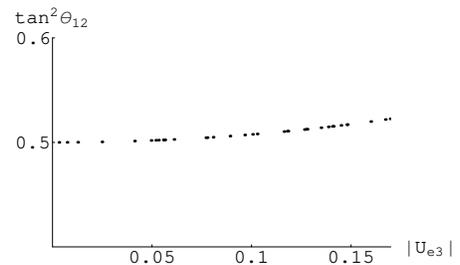}
\caption{ The allowed region on $|U_{e3}|-\tan^2\theta_{12} $ plane
  in the case of non-zero $\epsilon_1$.}
\end{center}
\end{wrapfigure}

As seen in  Eq.(\ref{MNS}),  $U_{e1}$ and $U_{e2}$ are independent of 
$\epsilon_1$  in the first order approximation.
 The effect of ${\cal O}(\epsilon_1^2)$ increases  $\tan^2\theta_{12}$
 toward the larger value  than the tri-maximal mixing $1/2$.
Taking the recent  experimental data as input \cite{PDG}:
\begin{eqnarray}
&&\Delta m_{32}^2=(1.9\sim 3.0)\times 10^{-3} 
{\rm eV}^2 \ ,
\quad  \sin^2 2\theta_{23}> 0.92   \nonumber \\
&&\Delta m_{21}^2= (8.0^{ +0.4}_{ -0.3}) \times 10^{-5} {\rm eV}^2 \ ,
\quad  \sin^2 2\theta_{12}=0.86^{+0.03}_{-0.04}  \ ,
\end{eqnarray}
\noindent
at $90\%$C.L.,
 we plot the allowed region on $|U_{e3}|-\tan^2\theta_{12} $ plane
in Figure 1, where we take 
 $a$, $b$, $d$  and $\epsilon_1$ as complex parameters.
 We also take the experimental upper bound
\begin{eqnarray}
\sin^2 2\theta_{\rm 13} < 0.19 \ .
\end{eqnarray}

The prediction of  $\tan^2\theta_{12}$ is almost
 at the high end of the experimental  allowed region,
while the  experimental central value is $\tan^2\theta_{12}=0.45$.
If we take  $\tan^2\theta_{12}=0.502$ as the upper bound,
 we get the bound of  $|U_{e3}|\leq 0.05$,
where $|\epsilon_1|$ is  smaller than $0.35$.
Therefore, the relation $b=c$ should be satisfied at the  $35\%$ level.
Therefore, it is preferred to decouple $\xi'$ and $\xi''$ 
in the Lagrangian of Eq.(\ref{La}), that is $b=c=0$.

It may be noticed that  direct  corrections to neutrino masses 
induce small values of  $b$ and $c$
even if  $\xi'$ and $\xi''$ are decoupled in the Yukawa couplings.
Since these corrections  on neutrino masses and mixing angles  are not
   leading ones, we neglect them hereafter.

\subsection{$a\not =0$, $b=c=0$, $e\not = 0$, $f\not =0$}

Let us consider the simple case of $b=c=0$.
Since the tri-bimaximal mixing with the relevant neutrino mass spectrum 
is realized at the limit of $\epsilon_2=\epsilon_3=0$  \cite{Ma20051},
  evolutions of the  non-zero  $\epsilon_2$ and $\epsilon_3$
 lead to  the deviation from the tri-bimaximal mixing. 

The  $\epsilon_2$ and $\epsilon_3$
 are expected to be real  if we suppose
    the vacuum expectation values  $(v_1', v_2', v_3')$ to be real.
 Since $a$ is taken to be real in general,
 we have a  phase $\phi_d$ in addition to the $A_4$ phase $\omega$.  
Then we take following parameters: 
\begin{eqnarray}
  a \ , \qquad  d=|d| \ e^{i\phi_d} \ , \qquad
 \epsilon_2\    \ , \qquad
\epsilon_3 \   \ ,
\end{eqnarray}
\noindent
where $a$,  $\epsilon_2$  and $\epsilon_3$  denote  real ones
\footnote{The specific case of $\epsilon_2=-\epsilon_3$ was investigated
  by Ma \cite{Ma20041}.  However, $b=c\not=0$ was assumed
 there, therefore, our numerical result does not  include his one.}.
  In the case  of  $\epsilon_2=\epsilon_3=0$,
 parameters $a, \ |d|, \ \phi_d$ are given in terms of
$\Delta m^2_{\rm atm}$ and $\Delta m^2_{\rm sol}$ as shown in Appendix.

As seen in  Eq.(\ref{MNS}), we have an approximate relation
\begin{eqnarray}
U_{\mu 3}\simeq -\frac{1}{\sqrt{2}}-U_{e3} \ .
\label{relation}
\end{eqnarray}
It is not easily to test this relation in the future experiments
 unless the phase of  $U_{e3}$ is known.

  It is noticed  in  Eq.(\ref{mass}) with $b=0$ that
 one  gets the neutrino mass spectrum with the normal mass hierarchy,
  but cannot get the inverted one \cite{Ma20051}.
We show the numerical results in Figure 2 and Table I.
 As seen in Fig.2(a), plots on  $\tan^2\theta_{12}-|U_{e3}|$ plane
 cover all experimental allowed region.
Therefore, there is no prediction in this case.
On the other hand,
 plots on  $\sin^22\theta_{23}-|U_{e3}|$ plane 
in Fig.2(b) indicate  the correlation between   $\sin^22\theta_{23}$
and $|U_{e3}|$.
One expects $\sin^22\theta_{23}\geq 0.98$ if  $|U_{e3}|\leq 0.05$
will be confirmed in the future experiments.
In the case of  $|U_{e3}|\leq 0.01$, 
  $\theta_{23}$   deviates    scarcely  from
the maximal mixing while   $\theta_{12}$ could  deviate considerably 
 from the  tri-maximal mixing.

\begin{figure}
\begin{center}
\includegraphics[width=7 cm]{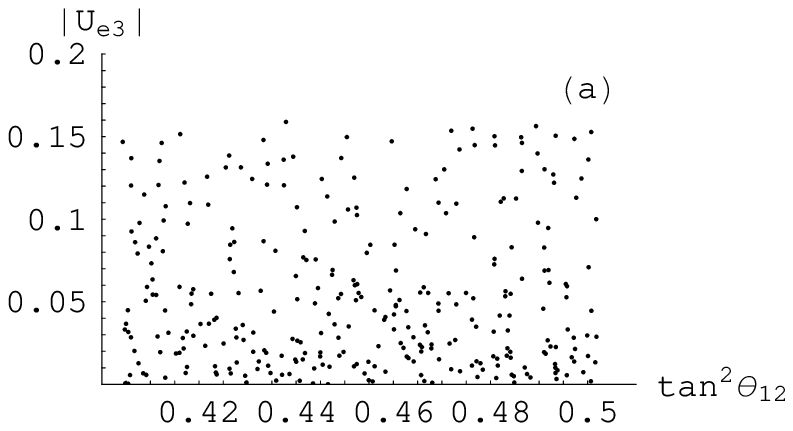} \qquad\qquad
\includegraphics[width=7 cm]{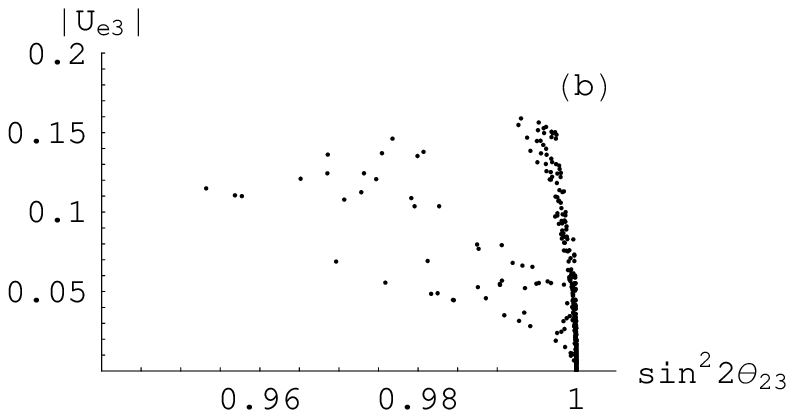}
\vskip 1 cm
\includegraphics[width=7 cm]{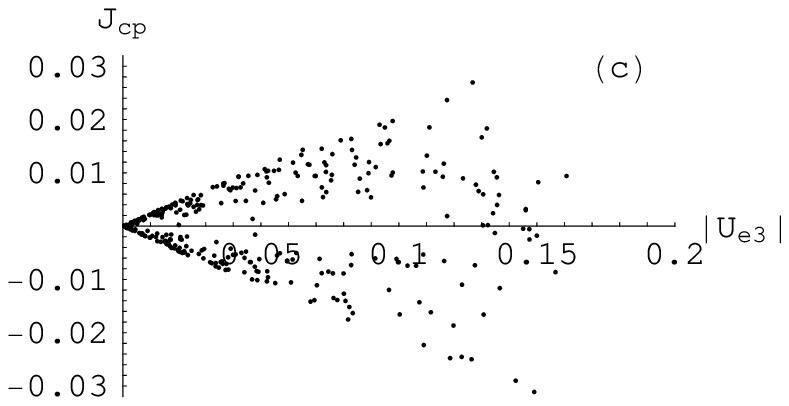} \qquad\qquad
\includegraphics[width=7 cm]{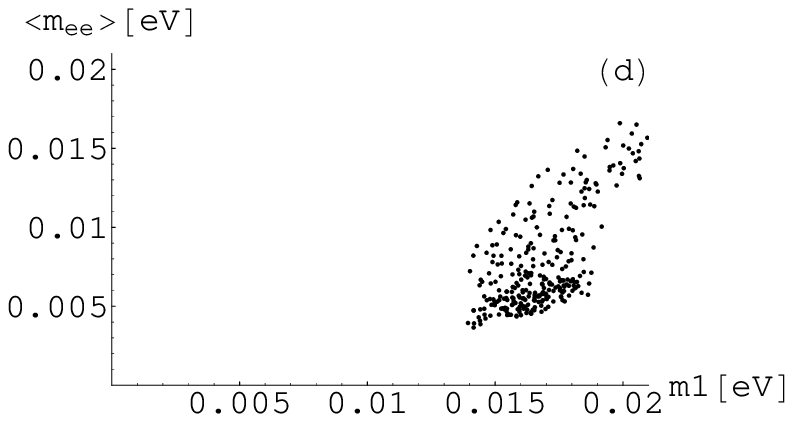}
\caption{ The allowed region on (a) $\tan^2\theta_{12}-|U_{e3}|$, 
(b) $\sin^22\theta_{23}-|U_{e3}|$, (c) $|U_{e3}|-J_{CP}$,  and
(d) $m_1-\langle m_{ee}\rangle$  planes
in the case of the normal  mass hierarchy of neutrinos with 
  non-zero $\epsilon_2$ and  $\epsilon_3$.}
\end{center}
\end{figure}

We plot allowed region 
 $J_{CP}$, which is the Jarlskog invariant \cite{Jcp} in Fig.2(c).
Since the CP violation comes from phases of  $\omega$ and $\phi_d$,
the predicted region is rather wide even if $|U_{e3}|$ is fixed.
The predicted absolute value reaches $0.03$.
\begin{figure}
\begin{center}
\includegraphics[width=6.5 cm]{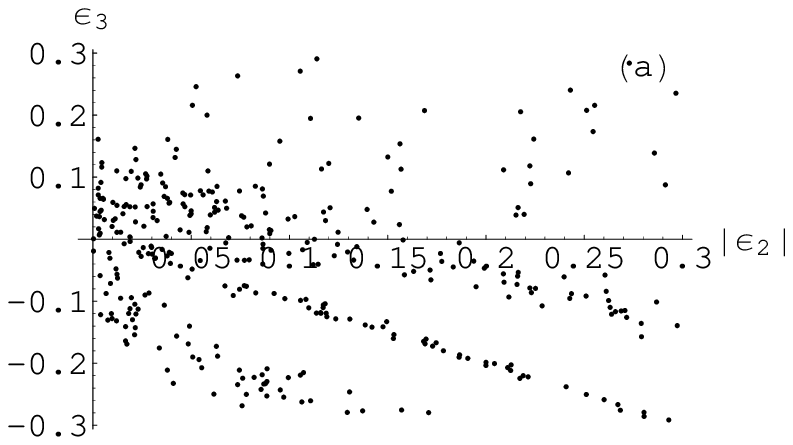} \qquad
\includegraphics[width=6.5  cm]{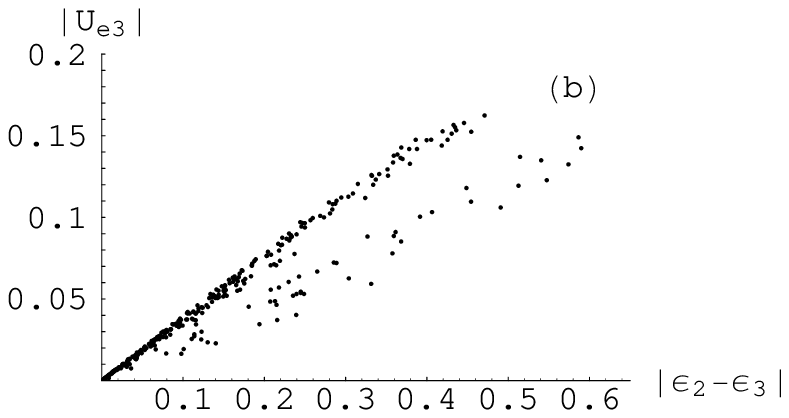}
\caption{ The allowed region on (a) $|\epsilon_2| - \epsilon_3 $ 
and  (b) $|\epsilon_2 - \epsilon_3| -|U_{e3}|$ planes.}
\end{center}
\end{figure}

In Fig.2(d), we show the predicted effective Majorana neutrino mass 
$\langle m_{ee} \rangle$, which is related with
the neutrinoless double beta decay rate, 
\begin{equation}
\langle m_{ee} \rangle=|M^f_{\nu}[1,1]|=\left|\ 
m_1c_{12}^2c_{13}^2e^{i \rho}+m_2s_{12}^2c_{13}^2e^{i\sigma}
+m_3s_{13}^2e^{-2i\delta_D} \ \right|\ ,
\end{equation}
\noindent
where $c_{ij}$ and  $s_{ij}$ denote
 $\cos \theta_{ij}$ and $\sin \theta_{ij}$, respectively, 
$\delta_D$ is a so called the Dirac phase, and $\rho,\sigma $ 
are the Majorana phases.
As seen in Fig.2(d),  
$\langle m_{ee} \rangle\geq 3.5\ {\rm meV}$ is predicted.
The magnitude increases proportional to the value of $m_1$
in the case of the  normal mass  hierarchy of  neutrinos.
The large value  $20\  {\rm meV}$ is expected  for  rather
degenerate neutrino masses.

\begin{wrapfigure}{r}{6.2cm}
\begin{center}
\includegraphics[width=6.0 cm]{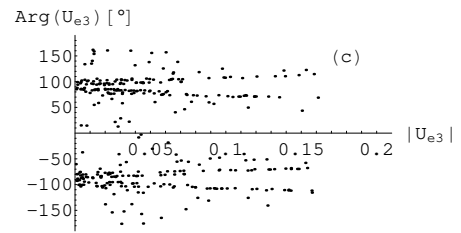}
\caption{ The predicted  phase of $U_{e3}$ versus $|U_{e3}|$
 in the case of non-zero $\epsilon_2$ and  $\epsilon_3$. 
The  phase  is predicted  around   $90^\circ$.}
\end{center}
\end{wrapfigure}
It may be useful to comment on the allowed values of
$\epsilon_2$ and $\epsilon_3$. 
 We show the allowed region on $|\epsilon_2| - \epsilon_3$ plane
 and  $|\epsilon_2 - \epsilon_3| -|U_{e3}|$ plane in Figure 3.
 Both  $\epsilon_2$ and $\epsilon_3$ are varied 
in the restricted region $-0.3\sim 0.3$ as seen in Fig.3(a).
It is found in Fig.3(b) that $|U_{e3}|$ is approximately proportional 
to the magnitude of  $\epsilon_2 - \epsilon_3$.
Since  $|\epsilon_2|$ and $|\epsilon_3|$ are cut at $0.3$ by hand,
$|U_{e3}|$ is bounded by $0.16$.

We also present the phase of $U_{e3}$, which is shown  in  Figure 4.
The  phase  of $U_{e3}$  is predicted  around   $90^\circ$,
which is  almost independent of  $|U_{e3}|$.
Therefore, $\sin^2 2\theta_{23}$    deviates at most 
  in a few percent  from the maximal mixing as seen in 
Eq.(\ref{relation}).

\subsection{Inverted mass hierarchy}
In this subsection, we discuss  the case of the inverted mass hierarchy,
which was presented in the case of 
$a=0$, $b = c\not= 0$  \cite{Ma20051}.
In this case,  $b = c$ may be accidental because
 the $A_4$ symmetry does not guarantee this relation.
Therefore, we discuss this case only in phenomenological interest
of the inverted neutrino mass hierarchy.
We take following parameters: 
\begin{eqnarray}
  b=c \ , \qquad  d=|d| \ e^{i\phi_d} \ , \qquad
 \epsilon_2\    \ , \qquad
\epsilon_3 \   \ ,
\end{eqnarray}
\noindent
where $b$, $c$,   $\epsilon_2$  and $\epsilon_3$  denote  real ones.
In the case  of  $\epsilon_2=\epsilon_3=0$,
 parameters $b, \ |d|, \ \phi_d$ are given in terms of
$\Delta m^2_{\rm atm}$ and $\Delta m^2_{\rm sol}$ as shown in Appendix.

%
%
\begin{table}
\caption{Predictions in subsections  B and C}
\begin{center}
\begin{tabular}{|l|l|l|}\hline
&
$b=c=0$\ ,\ Normal.  &  
    $a=0$\ ,\ Inverted.    \\  \hline \hline

\ $\tan ^2 \theta _{12}$  &  
\ $0.404\sim 0.502$  &
    \ $0.443\sim 0.502$                    \\  \hline
                                   
\ $\sin ^2 2\theta _{23}$  &  
\ $0.95\sim 1$  &  
   \ $0.99\sim 1$         \\  \hline

\ $|J_{cp}|$  &
\ $\leq 0.031$    &  
   \ $\leq 0.023$              \\  \hline

\ $\langle m_{ee} \rangle$  &
\ $\geq 3.5 \ {\rm meV}$  &  
   \ $14 \sim 22\ {\rm meV}$     \\  \hline

\ $ |U_{e3}|$  &  
\ $ \leq 0.16$  &  
\ $\leq  0.10$     \\  \hline                                            
\end{tabular}
\end{center}
\end{table}
\begin{figure}
\begin{center}
\includegraphics[width=7 cm]{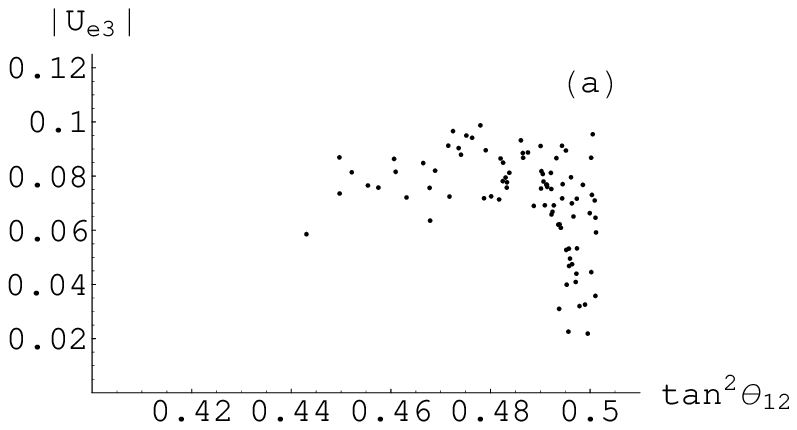} \qquad\qquad
\includegraphics[width=7 cm]{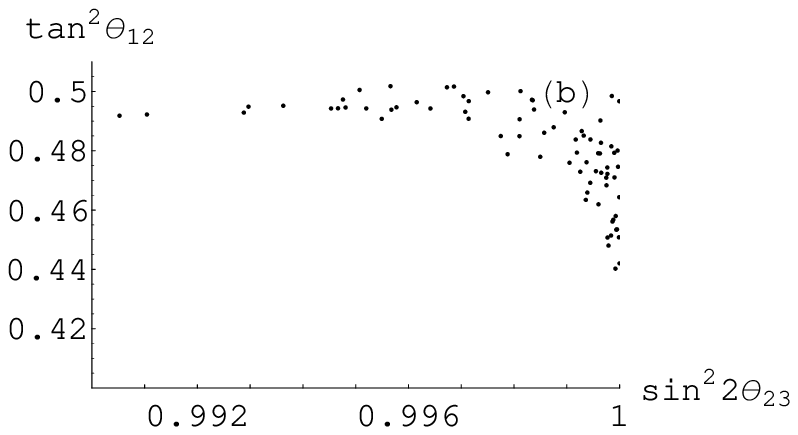}
\vskip 1 cm
\includegraphics[width=7 cm]{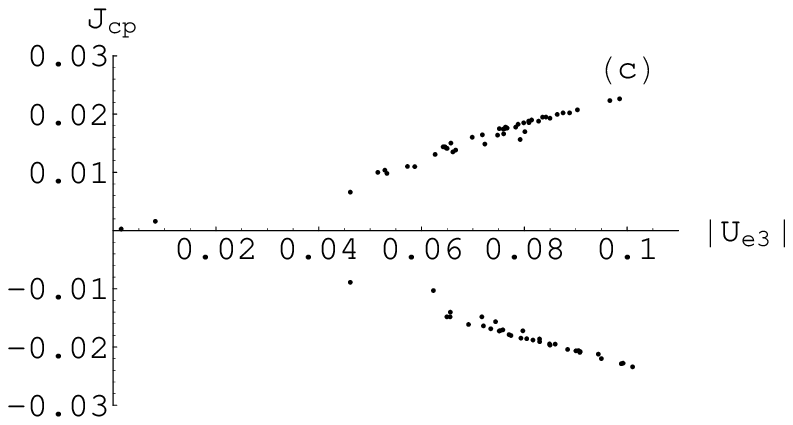} \qquad\qquad
\includegraphics[width=7 cm]{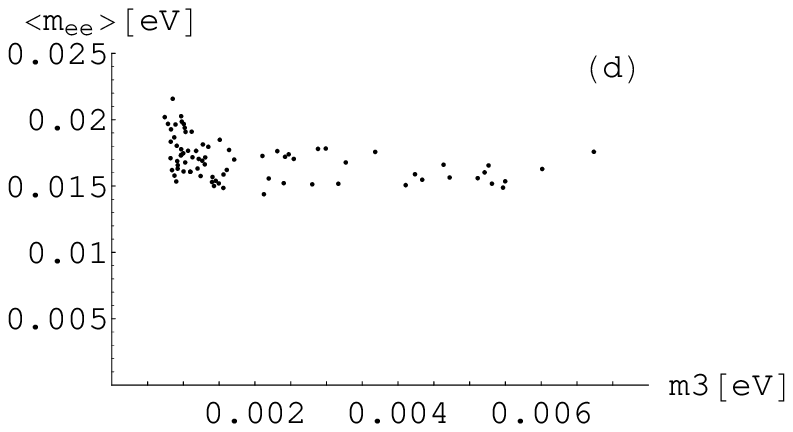}
\caption{ The allowed region on (a) $\tan^2\theta_{12}-|U_{e3}|$, 
(b) $\sin^22\theta_{23}-\tan^2\theta_{12}$, (c) $|U_{e3}|-J_{CP}$,  and
(d) $m_3-\langle m_{ee}\rangle$  planes
  in the case of the inverted mass hierarchy of neutrinos.}
\end{center}
\end{figure}

We show the numerical results in Figure 5 and Table I.
As $|U_{e3}|$ increases, the deviation from  $\tan^2{\theta_{12}}=1/2$
 can be  larger, while  $\sin^22\theta_{23}\geq 0.99$
 is obtained. As seen in Fig.5(b), the allowed region on  
 $\sin^22\theta_{23}-\tan^2{\theta_{12}}$ plane is restricted
  close to the tri-bimaximal mixing.
 It is noticed  that $|U_{e3}|$ is smaller than $0.10$. 

We also plot allowed region $J_{CP}$ in Fig.5(c).
It is remarkable that  $|J_{CP}|$ is almost determined if $|U_{e3}|$ is fixed.
This fact means that the inverted mass hierarchy is realized
 in the restricted value of   $\phi_d$.

The effective Majorana neutrino mass is
predicted to be a rather narrow range
$\langle m_{ee} \rangle=14\sim 22 \ {\rm meV}$ 
as seen in Fig.5(d).
Therefore, one may  expect that the neutrinoless double beta decay will be 
observed in the future experiments.

\subsection{The effect of the charged lepton} 
In this subsection, we discuss  the effect of the charged lepton mass matrix.
If the  vacuum alignment 
 $\langle \varphi \rangle =(v, v, v)$ is  sizeably corrected
 through  higher dimentional operators, we cannot neglect
the contribution from the charged lepton mass matrix 
to predict the deviation from the tri-bimaximal mixing.
In order to examine the effect of 
  correction parameters $\epsilon^{ch}_1,\ \epsilon^{ch}_2$
 in Eq.(\ref{epch}) clearly,
we take  $b = c=0$ and  $e=f= 0$ 
with the normal mass hierarchy of neutrinos  at first step.

We show numerical results in Figure 6 and Table II.
 As seen in Fig.6(a), plots  on  $\tan^2\theta_{12}-|U_{e3}|$ plane
 cover all experimental allowed region.
On the other hand, 
 plots on  $\sin^22\theta_{23}-\tan^2\theta_{12}$ plane 
in Fig.6(b) indicate $\sin^22\theta_{23}\geq 0.99$
while $\theta_{12}$ can  deviate from the  tri-maximal mixing considerably.
 
We also plot allowed region $J_{CP}$ in Fig.6(c).
Since the CP violation  is only due to   $\omega$, 
   $|J_{CP}|$ is clearly  determined if  $|U_{e3}|$ is fixed.
The predicted effective Majorana neutrino mass is
$\langle m_{ee} \rangle \geq 2.3 \ {\rm meV}$ 
as seen in Fig.6(d).

In order to see values  of $\epsilon_1^{ch}$ and  $\epsilon_2^{ch}$,
we plot allowed regions on 
 $\epsilon_1^{ch}-\epsilon_2^{ch}$  and 
 $|U_{e3}|-|\epsilon_1^{ch}-\epsilon_2^{ch}|$ planes. 
As seen in Fig.7(a),  the relative sign of
$\epsilon_1^{ch}$ and $\epsilon_2^{ch}$  is almost opossite,
and these  magnitudes are at most $0.3$.
Furtheremore, it is found that
 $|U_{e3}|$ is proportional to
 the magnitude of $\epsilon_1^{ch}-\epsilon_2^{ch}$ as seen in Fig.7(b).

We also examine the case of $a=0$, $b = c\not =0$ and $e=f=0$
with the inverted hierarchy of  neutrino masses.
The numerical result is not so changed in the above case
 as summarized in Table II.

 At the next step,  we present the numerical result
 in the case that 
 both charged lepton  and the neutrino mass matrices  contribute to
 the deviation from the tri-bimaximal mixing.
 Taking $|\epsilon_1^{ch}|\leq 0.05$,  $|\epsilon_2^{ch}|\leq 0.05$,
 $|\epsilon_2|\leq 0.05$, and  $|\epsilon_3|\leq 0.05$, which 
 guarantees that  
corrections of  higher-dimensional operators do not spoil
the leading order picture as discussed  in the previous work \cite{Alta2},
we predict the following values for the normal neutrino mass  hierarchy:
\begin{eqnarray}
&&\tan ^2 \theta _{12}=0.404\sim 0.502 \ , \quad
\sin ^2 2\theta _{23}=0.997\sim 1 \ , \quad 
|U_{e3}|\leq 0.047 \ , \nonumber\\ 
&& |J_{CP}|\leq 0.011\ ,  \quad
\langle m_{ee}\rangle \geq  4.2 \ {\rm meV} \ .
\end{eqnarray}
These predicted  mixing angles should be taken as a typical deviation 
from the tri-bimaximal mixing in the $A_4$ model.
\begin{figure}
\begin{center}
\includegraphics[width=6 cm]{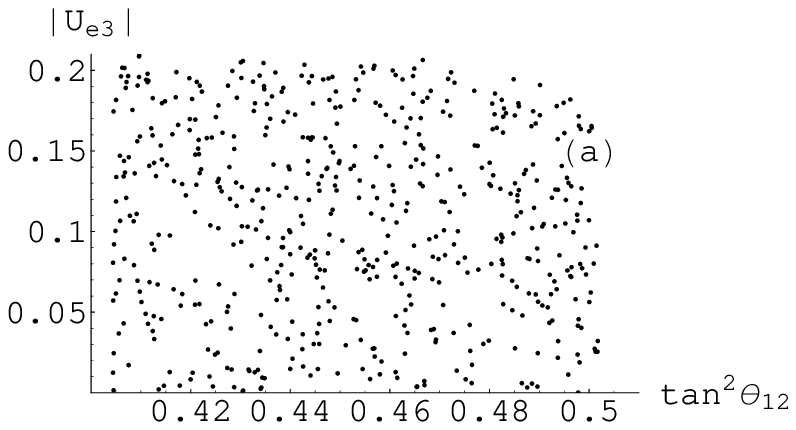} \qquad\qquad
\includegraphics[width=6 cm]{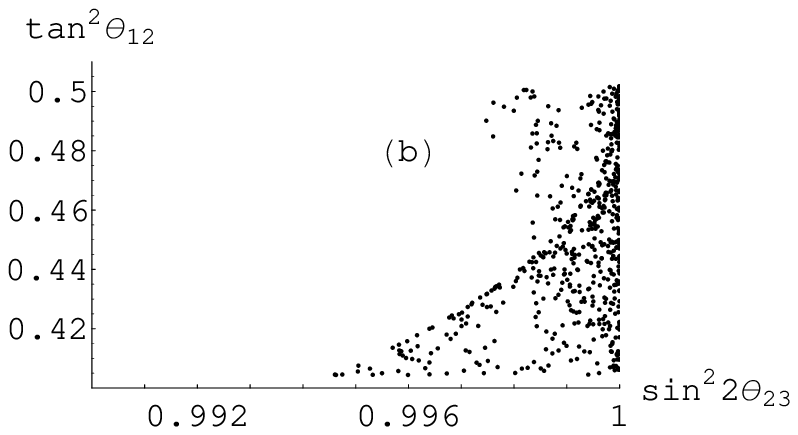}
\vskip 1 cm
\includegraphics[width=6 cm]{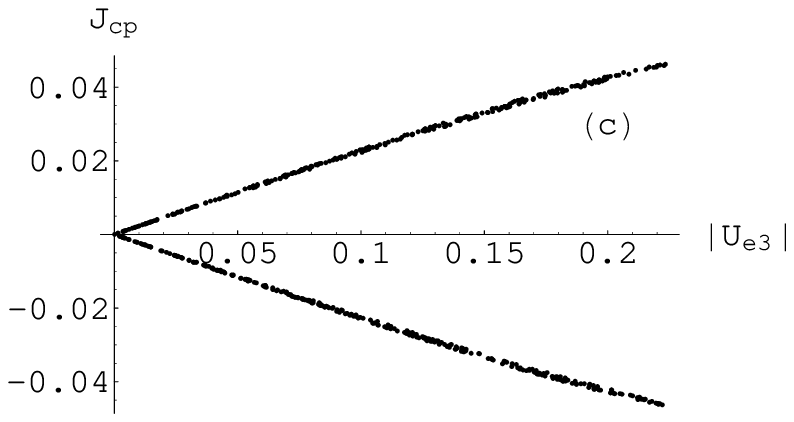} \qquad\qquad
\includegraphics[width=6 cm]{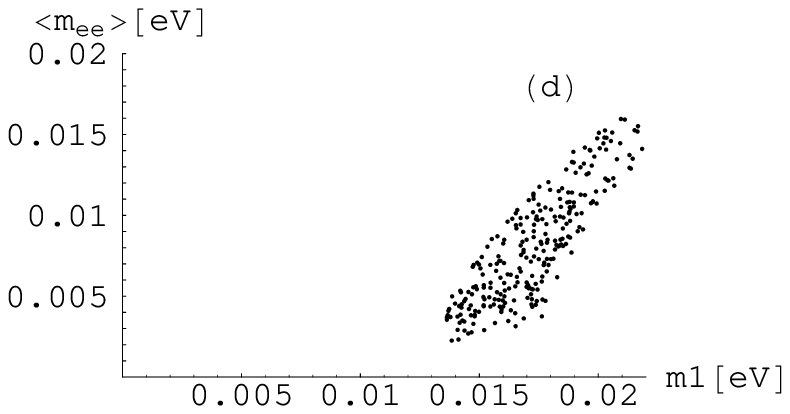}
\caption{ The allowed region on (a) $\tan^2\theta_{12}-|U_{e3}|$, 
(b) $\sin^22\theta_{23}-\tan^2\theta_{12}$, (c) $|U_{e3}|-J_{CP}$,  and
(d) $m_1-\langle m_{ee}\rangle$  planes
   with 
non-zero $\epsilon_1^{ch}, \epsilon_2^{ch}$  in the charged lepton sector.}
\end{center}
\end{figure}

\begin{figure}
\begin{center}
\includegraphics[width=6 cm]{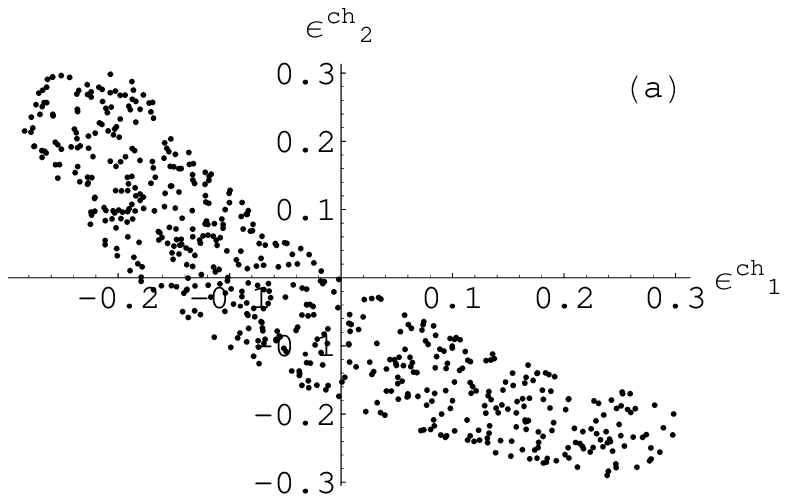} \qquad\qquad
\includegraphics[width=6 cm]{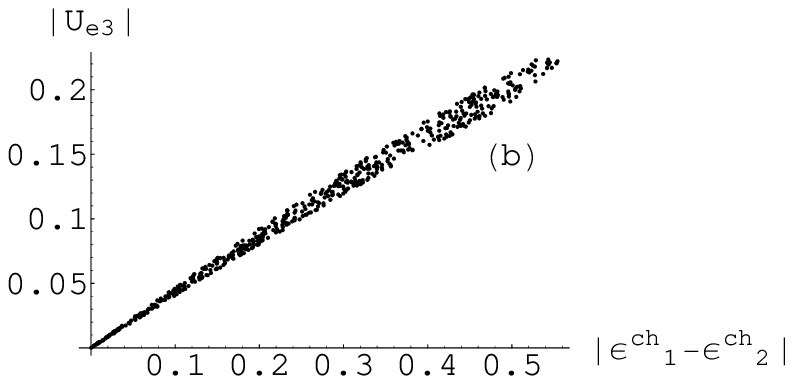}
\caption{ The allowed region on (a) $\epsilon^{ch}_1 - \epsilon^{ch}_2 $ 
and  (b) $|\epsilon^{ch}_1 - \epsilon^{ch}_2| -|U_{e3}|$ planes.}
\end{center}
\end{figure}
%
%

\begin{table}
\caption{Predictions in subsection D}
\begin{center}
\begin{tabular}{|l|l|l|}\hline

&
$b=0$\ ,\ Normal.  &  
    $a=0$\ ,\ Inverted.    \\  \hline \hline

\ $\tan ^2 \theta _{12}$  &  
\ $0.404\sim 0.502$  &
    \ $0.404\sim 0.502$                    \\  \hline
                                   
\ $\sin ^2 2\theta _{23}$  &  
\ $0.994\sim 1$  &  
   \ $0.995\sim 1$         \\  \hline

\ $|J_{cp}|$  &
\ $\leq  0.046$    &  
   \ $\leq  0.046$              \\  \hline

\ $\langle m_{ee} \rangle $ &
\ $\geq 2.3 \ {\rm meV}$  &  
   \ $14\sim 22\ {\rm meV}$     \\  \hline

 \ $|U_{e3}|$  &  
 \ $\leq 0.22$  &  
  \ $\leq 0.22$     \\  \hline                                            
\end{tabular}
\end{center}
\end{table}

\section{Summary}
 We have examined
the deviation from the tri-bimaximal mixing of the neutrino flavors
 in the framework of  the $A_4$ model.
Taking account   corrections of the vacuum alignment of flavon fields,
the deviation  from the tri-bimaximal mixing are estimated
quantitatively.

In the case of the normal mass hierarchy of   neutrinos,
there is   the correlation between   $\sin^22\theta_{23}$
and $|U_{e3}|$.
We  expect $\sin^22\theta_{23}\geq 0.98$ if  $|U_{e3}|\leq 0.05$
will be confirmed in the future experiments.
If the  stronger  bound  $|U_{e3}|\leq 0.01$ will be obtained in the future, 
  $\theta_{23}$ is expected to be almost maximal mixing while   $\theta_{12}$ could be deviated considerably from the tri-maximal mixing.

In the case of the inverted mass  hierarchy of   neutrinos,
 the deviation from  $\tan^2{\theta_{12}}=1/2$
 becomes larger  as   $|U_{e3}|$ increases.
The allowed region on  
 $\sin^22\theta_{23}-\tan^2{\theta_{12}}$ plane is restricted
  close to the tri-bimaximal mixing.
It is remarkable that  $|J_{CP}|$ is almost determined if $|U_{e3}|$ is fixed.
Moreover,
 the neutrinoless double beta decay is expected to  be observed
 in the future experiments
because the predicted  effective Majorana neutrino mass is
$\langle m_{ee} \rangle=14\sim 22 \ {\rm meV}$. 

 The deviation through the charged lepton sector is also  examined.
If   parameters are constrained such as 
 $|\epsilon_1^{ch}|\leq 0.05$,  $|\epsilon_2^{ch}|\leq 0.05$,
 $|\epsilon_2|\leq 0.05$, and  $|\epsilon_3|\leq 0.05$,
  $\tan^2{\theta_{12}}$ can deviate 
considerably  from the tri-maximal mixing  $1/2$
while 
$\sin ^2 2\theta _{23}\geq 0.997$ and  $|U_{e3}|\leq 0.047$ are obtained.
These values indicate a typical deviation 
from the tri-bimaximal mixing in the $A_4$ model.

 The precision measurements of 
the deviation from the tri-bimaximal mixing of the neutrino flavors
 provide  a crucial test of the $A_4$ flavor symmetry 
with  the vacuum alignment of flavon fields.

\acknowledgments{
The work of M.T. has been  supported by the
Grant-in-Aid for Science Research
of the Ministry of Education, Science, and Culture of Japan
Nos. 17540243 and 19034002.}

 \section*{Appendix}
 
In this appendix, we show relations 
 among mass eiganvalues and parameters 
 at the limit of  the  tri-bimaximal mixing.

In the case of $b=c=0$ with the  normal hierarchy,  mass eigenvalues $m_i^0$
 are given as
\begin{eqnarray}
m_1^0=a+|d| e^{i\phi_d}\ ,\quad m_2^0=a\ ,\quad m_3^0=-a+ |d| e^{i\phi_d}\ ,
\end{eqnarray}
\noindent where $a$ is taken to be real. Then, we have 
\begin{eqnarray}
\Delta m_{\rm atm}^2\equiv |m_3^0|^2- |m_1^0|^2=-4a|d| \cos{\phi_d}\ , \qquad
\Delta m_{\rm sol}^2\equiv |m_2^0|^2- |m_1^0|^2
=\frac{1}{2} \Delta m_{\rm atm}^2-|d|^2\ ,
\end{eqnarray}
\noindent
which give
\begin{eqnarray}
a=-\frac{\Delta m_{\rm atm}^2}{4|d| \cos{\phi_d}}\ , \qquad 
|d|=\sqrt{\frac{1}{2}\Delta m_{\rm atm}^2-\Delta m_{\rm sol}^2} \ .
\end{eqnarray}
%
%
%
 
In the case of  $a=0$ and $b=c\not= 0$ with  the inverted  hierarchy,
mass eigenvalues $m_i^0$ are given as
\begin{eqnarray}
m_1^0=-b+|d| e^{i\phi_d}\ ,\quad m_2^0=2b\ ,\quad  m_3^0=b+|d| e^{i\phi_d}\ ,
\end{eqnarray}
  where $b$ is taken to be real. Then, we have 
\begin{eqnarray}
\Delta m_{\rm atm}^2\equiv |m_1^0|^2- |m_3^0|^2=-4b|d| \cos{\phi_d}\ ,\quad 
\Delta m_{\rm sol}^2\equiv |m_2^0|^2- |m_1^0|^2=3b^2+2b|d| \cos{\phi_d}-|d|^2\ ,
\end{eqnarray}
\noindent which gives  
\begin{eqnarray}
b&=&-\frac{\Delta m_{\rm atm}^2}{4|d| \cos{\phi_d}},\\
\nonumber\\
|d|^2&=&-\frac{\Delta m_{\rm atm}^2}{4}-\frac{\Delta m^2_{\rm sol}}{2}
            +\frac{1}{4}\sqrt{(1+\frac{3}{\cos^2{\phi_d}})
                              \Delta m_{\rm atm}^4
                                  +4 \Delta m_{\rm atm}^2\Delta m_{\rm sol}^2
                                  +4 \Delta m_{\rm sol}^4
                                   }\ .
\end{eqnarray}



\begin{thebibliography}{99}
\bibitem{maltoni}
M.~Maltoni, T.~Schwetz, M.~Tortola, and J.W.F.~Valle,
New J.~Phys. {\bf 6},  122  (2004).

\bibitem{fogli}
G.L.~Fogli, E.~Lisi, A.~Marrone,  and A.~Palazzo, 
Prog. Part. Nucl. Phys. {\bf 57}, 742 (2006).

\bibitem{HPS}
P.F. Harrison, D.H. Perkins, and W.G. Scott, 
Phys. Lett. B  {\bf 530}, 167 (2002);\\
P.F. Harrison and W.G. Scott, 
Phys. Lett. B {\bf 535}, 163 (2002).
 
\bibitem{S3}
S.~Pakvasa  and H.~Sugawara,
Phys.\ Lett.\ B {\bf 73}, 61 (1978);
\\
D.~Wyler, Phys.\ Rev.\ D {\bf 19}, 330 (1979);
\\
L.J.~Hall, and H.~Murayama,
Phys.\ Rev.\ Lett.\ {\bf 75}, 3985 (1995);
\\
R.~Dermisek and S.~Raby,
Phys.\ Rev.\ D {\bf 62},  015007 (2000) ;
\\
R.N.~Mohapatra, A.~Perez-Lorenzana, and C.A.~de Sousa Pires,
Phys.\ Lett.\ B {\bf 474},  355 (2000);
\\
J.~Kubo, A.~Mondragon, M.~Mondragon, and E.~Rodriguez-Jauregui,
Prog.\ Theor.\ Phys.\ {\bf 109}, 795 (2003),
[Erratum-ibid.\ {\bf 114}, 287 (2005)];
\\
T.~Kobayashi, J.~Kubo,  and H.~Terao,
Phys.\ Lett.\ B {\bf 568}, 83 (2003);
\\
P.F.~Harrison  and W.G.~Scott,
Phys.\ Lett.\ B {\bf 557}, 76 (2003);
\\
J.~Kubo, H.~Okada,  and F.~Sakamaki,
Phys.\ Rev.\ D {\bf 70},  036007 (2004);
\\
S.-L.~Chen, M.~Frigerio, and E.~Ma,
Phys.\ Rev.\ D {\bf 70},  073008 (2004);
[Erratum-ibid.\ D {\bf 70},  079905  (2004)];
\\
W.~Grimus and L.~Lavoura, 
JHEP {\bf 0508}, 013 (2005);
\\
Y.~Koide, Phys.\ Rev.\ D {\bf 73}, 057901 (2006);
\\
R.N. Mohapatra, S. Nasri, and Hai-Bo,  Phys. Lett. {\bf B639},  318 (2006);
\\
N.~Haba and K.~Yoshioka,
Nucl.\ Phys.\ B {\bf 739}, 254 (2006);
\\
S. Kaneko, H. Sawanaka, T. Shingai,  M. Tanimoto, and K. Yoshioka,
Prog. Theor. Phys. {\bf 117}, 161 (2007). 

\bibitem{Sn}
Y.~Yamanaka, H.~Sugawara, and S.~Pakvasa,
Phys.\ Rev.\ D {\bf 25}, 1895 (1982); [Erratum-ibid. D {\bf 29},
2135 (1984)];
\\
C.~Hagedorn, M.~Lindner,  and R.~N.~Mohapatra, 
JHEP {\bf 0606}, 042 (2006).

\bibitem{D4}
W.~Grimus and L.~Lavoura,
Phys.\ Lett.\ B {\bf 572},  189 (2003);
\\
W.~Grimus, A.~S.~Joshipura, S.~Kaneko, L.~Lavoura, and M.~Tanimoto,
JHEP {\bf 0407}, 078 (2004);
\\
W.~Grimus, A.~S.~Joshipura, S.~Kaneko, L.~Lavoura, H.~Sawanaka,
and M.~Tanimoto,
Nucl.\ Phys.\ B {\bf 713}, 151 (2005);
\\
P. Ko, T. Kobayashi, J. Park, and S. Raby, hep-ph/0704.2807;
\\
A. Blum, C. Hagedron, and M. Lindner, hep-ph/0709.3450.
\bibitem{Q6}
K.S.~Babu and J.~Kubo, Phys.\ Rev.\ D {\bf 71},  056006 (2005);
\\
Y.~Kajiyama, E.~Itou and J.~Kubo,
Nucl.\ Phys.\ B {\bf 743}, 74 (2006).

\bibitem{Q8}
M.~Frigerio, S.~Kaneko, E.~Ma,  and M.~Tanimoto,
Phys.\ Rev.\ D {\bf 71},  011901 (2005).

\bibitem{others}
E.~Derman and H.S.~Tsao,
Phys.\ Rev.\ D {\bf 20}, 1207 (1979);
\\
D.~Chang, W.Y.~Keung, and G.~Senjanovic,
Phys.\ Rev.\ D {\bf 42}, 1599 (1990);
\\
D.B.~Kaplan  and M.~Schmaltz,
Phys.\ Rev.\ D {\bf 49}, 3741 (1994);
\\
P.H.~Frampton and T.W.~Kephart,
Int.\ J.\ Mod.\ Phys.\ A {\bf 10}, 4689  (1995);
\\
A.~Aranda, C.D.~Carone, and R.F.~Lebed,
Phys.\ Lett.\ B {\bf 474}, 170 (2000);
\\
N.~Haba, A.~Watanabe, and K.~Yoshioka,
Phys.\ Rev.\ Lett.\ {\bf 97},  041601 (2006);
\\
C.~Hagedorn, M.~Lindner, and F.~Plentinger,
Phys.\ Rev.\ D {\bf 74},  025007 (2006);
\\
F.~Feruglio and Y. Lin, hep-ph/0712.1528;
\\
  P.~H.~Frampton and T.~W.~Kephart,
  Int.\ J.\ Mod.\ Phys.\  A {\bf 10}, 4689 (1995);
\\
  A. Aranda, C.D. Carone, and R.F. Lebed,
  Int.\ J.\ Mod.\ Phys.\  A {\bf 16S1C}, 896 (2001);
  Phys.\ Rev.\  D {\bf 62},  016009 (2000);
  Phys.\ Lett.\  B {\bf 474}, 170 (2000);
\\
 P.~H.~Frampton and T.~W.~Kephart,
  JHEP {\bf 0709}, 110 (2007);
\\
 P.~D.~Carr  and P.~H.~Frampton, hep-ph/0701034;
\\
H. Frampton and  S. Matsuzaki,
hep-ph/0712.1544;  hep-ph/0710.5928;
\\
I. de Medeiros Varzielas, S.F. King, and G.G. Ross,
 Phys. Lett. B {\bf 648}, 201 (2007);
\\
M-C. Chen and K.T. Mahanthappa, Phys. Lett. B {\bf 652}, 34 (2007);
\\
C. Luhn, S. Nasri, and P. Ramond, Phys. Lett. B {\bf 652},  27 (2007). 

\bibitem{A4}
E.~Ma and G.~Rajasekaran,
Phys.\ Rev.\ D {\bf 64},  113012 (2001);
\\
E.~Ma,
Mod.\ Phys.\ Lett.\ A {\bf 17}, 2361 (2002);
\\
K.S.~Babu, E.~Ma,  and J.W.F.~Valle,
Phys.\ Lett.\ B {\bf 552}, 207 (2003).

\bibitem{A41}
M.~Hirsch, J.C.~Romao, S.~Skadhauge, J.W.F.~Valle, and A.~Villanova del
Moral,
Phys.\ Rev.\ D {\bf 69},  093006 (2004);
\\
A.~Zee,
Phys.\ Lett.\ B {\bf 630}, 58 (2005);
\\
X.-G.~He, Y.-Y.~Keum,  and R.R.~Volkas,
JHEP {\bf 0604}, 039 (2006);
\\
E.~Ma, H.~Sawanaka, and M.~Tanimoto, Phys. Lett. B {\bf 641}, 
 301 (2006);
\\
S. F. King and  M.  Malinsk\'y, Phys. Lett. B {\bf 645}, 351 (2007);
\\
G. Altarelli, F. Feruglio, and Yin Lin, Nucl. Phys. B {\bf 775}, 31 (2007); 
\\
M. Hirsch, A. S. Joshipura, S. Kaneko, and J.W.F. Valle,
Phys. Rev. Lett. {\bf 99},   151802 (2007);
\\
F. Bazzocchi, S. Kaneko, and S. Morisi, hep-ph/0707.3032;
\\
L. Lavoura and H. K\"uhb\"ock,  
Mod. Phys. Lett. A {\bf 22}, 181 (2007); hep-ph/0711.0670;
\\
B. Adhikary, B. Brahmachari, Ambar Ghosal,  Ernest Ma, and M. K. Parida,
Phys. Lett. B {\bf 638}, 345 (2006); 
\\
B. Adhikary and Ambar Ghosal,  Phys. Rev. D  {\bf 75}, 073020 (2007).

 \bibitem{Ma20051}
 E. Ma, Phys.\ Rev. D {\bf 72},  037301 (2005).

 \bibitem{Ma20041}
 E. Ma, Phys.\ Rev. D {\bf 70},  031901 (2004).

 \bibitem{Alta1}
 G. Altarelli and F. Feruglio, Nucl. Phys. B {\bf 720}, 64  (2005). 

 \bibitem{Alta2}
 G. Altarelli and F. Feruglio, Nucl. Phys. B {\bf 741}, 215 (2006).

\bibitem{MNS}
Z. Maki, M. Nakagawa,  and S. Sakata,
Prog. Theor. Phys. {\bf 28}, 870 (1962).

\bibitem{PDG}
 Particle Data Group, http://pdg.lbl.gov/ (2007).

\bibitem{Jcp}
C. Jarlskog, Phys. Rev. Lett. {\bf 55}, 1039 (1985).

\end{thebibliography}
\end{document}